\documentclass[]{spie}
\usepackage[english]{babel}
\usepackage[utf8x]{inputenc}
\usepackage[T1]{fontenc}
\usepackage{amsmath}
\usepackage{gensymb}
\usepackage{amsfonts}
\usepackage{chngpage}
\usepackage{graphicx}
\usepackage{float}
\usepackage{authblk}
\usepackage[colorlinks=true, allcolors=blue]{hyperref}
\usepackage[colorinlistoftodos]{todonotes}
\usepackage{nopageno}
\title{Characterization of lemniscate atmospheric aberrations in Gemini Planet Imager data }
\author[1]{Alexander Madurowicz}
\author[1]{Bruce A. Macintosh}
\author[1]{Jean-Baptiste Ruffio}
\author[1]{Jeffery Chilcote}
\author[2]{Vanessa P. Bailey}
\author[3]{Lisa Poyneer}
\author[1]{Eric Nielsen}
\author[1]{Andrew P. Norton}
\affil[1]{Kavli Institute for Particle Astrophysics and Cosmology, Stanford University, Stanford, CA, 94305}
\affil[2]{Jet Propulsion Laboratory, California Institute of Technology, Pasadena, CA, 91109}
\affil[3]{Lawrence Livermore National Laboratory,  Livermore, CA 94550}
\authorinfo{Send correspondence to A. Madurowicz E-mail: amaduro@stanford.edu}
\begin{document}
\maketitle
\begin{abstract}
A semi analytic framework for simulating the effects of atmospheric seeing in Adaptive Optics systems on an 8-m telescope is developed with the intention of understanding the origin of the wind-butterfly, a characteristic two-lobed halo in the PSF of AO imaging. Simulations show that errors in the compensated phase on the aperture due to servo-lag have preferential direction orthogonal to the direction of wind propagation which, when Fourier Transformed into the image plane, appear with their characteristic lemniscate shape along the wind direction. We develop a metric to quantify the effect of this aberration with the fractional standard deviation in an annulus centered around the PSF, and use telescope pointing to correlate this effect with data from an atmospheric models, the NOAA GFS. Our results show that the jet stream at altitudes of 100-200 hPa (equivalently 10-15 km above sea level) is highly correlated (13.2$\sigma$) with the strong butterfly, while the ground wind and other layers are more or less uncorrelated.

\end{abstract}
\section{Introduction}
The Gemini Planet Imager (GPI) is an instrument installed on the Gemini South Telescope in Cerro Pachon, Chile, designed to search for thermal emission from young hot extrasolar planets at wide angular separation\cite{Macintosh2014}. The GPI Exoplanet Survey (GPIES) is well under way, with many successful detections\cite{Macintosh2015}, but contrast remains limited by residual atmospheric aberrations\cite{Ruffio2017}\cite{Poyneer2016} and imperfections in the Adaptive Optics (AO) system\cite{MalesGuyon17}. Notably, the presence of the wind butterfly (so-called because of its figure-8 or lemniscate shape) scatters significant light into the coronagraphic dark hole and causes the residual point-spread function (PSF) to break azimuthal symmetry in the image plane, which reduces the final contrast ratio. This PSF pattern is consistent with wavefront errors from servo lag\cite{Rigaut1998}. A $\sim$$2$ ms delay in positioning of the deformable mirror in response to the wavefront sensor causes a displacement between the actual phase errors on the aperture from atmospheric turbulence when compared to the phase of the applied correction. The Fourier transform of this particular pattern has preferential direction, and appears in the image plane as the wind butterfly. 

In this paper, we will demonstrate the servo-lag error as the origin of the wind butterfly in simulations of AO systems based on a theoretical turbulence model that is well-verified with a Kolmogorov power spectrum, as well as presenting our results in analyzing the appearances of the wind butterfly in the GPI Exoplanet Survey data. We have developed a metric to identify image subsets where this effect is highly apparent. The fractional standard deviation in an annulus centered about the PSF is zero in the case in azimuthal symmetry, and approaches one for cases of extreme azimuthal asymmetry. Using a chi-squared minimization routine with a model of the image's azimuthal dependence, we can extract the wind direction in the image plane, which, along with telescope pointing information, identifies the wind vector in three dimensions. This allows us explore correlations between the wind vector as pointed by the butterfly, and the direction of high-altitude winds from the NOAA GFS database.  Evaluating the azimuthal variations in planet detectability caused by the butterfly will help us to evaluate survey sensitivity and potential improvements from faster AO correction.

\section{Butterflies in Simulations}
In order to motivate understanding the appearance of the wind-butterfly in the GPIES, we would like to first develop a semi-analytic framework that describes turbulence in the atmosphere and how adaptive optics respond to provide a clear picture of the mechanism that brings the wind butterfly into existence. We begin with a short review on standard models of turbulence and their effect on optical imaging systems. 

\subsection{Theoretical Turbulence Models and Adaptive Optics Simulations}
Tartarski\cite{Tatarski1961} has shown that the fluctuations in the optical index of refraction in three dimensions for a Kolmogorov turbulence spectrum follow the form
\begin{equation}
\Phi_N(\kappa, z) = 0.033 C_N^2(z)\kappa^{-11/3}
\end{equation}
Where $C_N^2$ is the index of refraction structure constant and and $\kappa = 2\pi/l$ is the spatial wavevector for an eddy of size $l$. Here, we use the standard Kolmogorov power spectrum, which is fractally self-similar at all length scales, although it is in principle simple to extend this model to a Von-Karman spectrum by attenuating the power above and below the outer and inner scales. From the square root of the power spectrum, we can find the fluctuations from the inverse Fourier Transform according to Johansson\cite{Johansson1994} with

\begin{equation}
\delta N (\vec{x},z) = \textrm{Re}\Big[\mathcal{F}^{-1}\Big(\xi(\vec{\kappa},z)\sqrt{\Phi_N(\kappa,z)}\Big)\Big]
\end{equation}
where $\delta N$ are the fluctuations of the index of refraction from unity in parts per million, $\xi$ is a zero-mean unit-variance complex hermitian Gaussian noise process, and $\mathcal{F}^{-1}$ is the unnormalized inverse Discrete Fourier Transform (DFT) given by
\begin{equation}
\eta_{\textrm{ab}} = \mathcal{F}^{-1}(\tilde{\eta}_{\textrm{pq}}) = \sum_{p=0}^{P-1} \sum_{q=0}^{Q-1} \tilde{\eta}_{\textrm{pq}} \exp\Big[2\pi i\Big(\frac{pa}{P} + \frac{qb}{Q}\Big)\Big]
\end{equation}
for a discrete array of size $P\times Q$ with $P,Q \in \mathbb{N}$. The discrete indices $p,a \in {0,1, ..., P-1}$ and $q,b \in {0,1, ..., Q-1}$ exist in Fourier and configuration space, respectively. The corresponding forward Fourier Transform simply includes negation in the exponent, and we have to pay careful attention the the normalization factor used by a routine such as np.fft.fft2, which includes a normalization of $\frac{1}{PQ}$ on the inverse transform, but no normalization on the forward transform by default.

The optical path length of a wavefront traversing a turbulent layer in the atmosphere from zenith can be found to first order by integrating the index of refraction over the thickness of the layer, and the accumulated phase is simply the wavevector of the ray $k = 2\pi/\lambda$ times the optical path length.
\begin{equation}
\phi_i(\vec{x}) = k\int_{z_i}^{z_i+\Delta z_i} n(\vec{x},z) \textrm{d}z = k n(\vec{x},z)\Delta z_i
\end{equation}
Here $\vec{x} = (x, y)$ is the coordinate system in the aperture at $z=0$, $\Delta z_i$ is the range of altitudes relevant to the turbulent layer at altitude $z_i$, and the baseline index of refraction of the atmosphere can be approximated\cite{Hardy1998} with 
\begin{equation}
N \equiv (n-1)10^6 \approx 77.6 \frac{P}{T}
\end{equation}
where $P$ and $T$ are the pressure (in millibars or equivalently hPa) and temperature (in Kelvin) of the atmosphere for a particular altitude. We will obtain a model of the index of refraction for the baseline atmosphere in a later section, but once that is acquired we simply add the fluctuations on top $N + \delta N$ to simulate a particular turbulent instance. For non-zenith observations an additional term of $\sec{\zeta}$ where $\zeta$ is the zenith angle should be included in the integral in (4) to account for additional atmospheric depth. When the accumulated phase on the aperture is very large, we can subtract off the average phase, which is equivalent to removing the piston term from a Zernike Polynomial.\cite{MalesGuyon17}

Furthermore, we assume the Taylor frozen-flow hypothesis, which requires that the timescale for turbulence is much greater than the time delay with which the AO system will respond. For our simulation, this means that the fluctuations in the field of view simply propagate by translations due to the wind velocity, which can be expressed by
\begin{equation}
\delta N(\vec{x}+\vec{v}(z)\tau,t_0+\tau) = \delta N(\vec{x}, t_0)
\end{equation}
where $\vec{v}(z)$ is the wind velocity at altitude $z$, which is assumed to lie only in the plane at altitude with no vertical component, $t_0$ is an particular instant in time, and $\tau$ is the total time delay for the adaptive optics system to respond to a measurement from the wavefront sensor. We also assume a perfect noiseless wavefront sensor and deformable mirror whose only flaw is a delayed response to test our hypothesis for the origin of the wind butterfly. In essence, this is an ideal open-loop AO simulation. The expression for the compensated phase in the aperture is then
\begin{equation}
\phi_c = \sum_{i=1}^L [\phi_i(\vec{x},t_0) - \phi_i(\vec{x},t_0-\tau)]
\end{equation}
where we sum over the contributions from $L$ turbulent layers at altitudes $z_i$ with a flat step interpolation scheme for the structure constant. For a continuous wind velocity profile, one could transform this summation into an appropriate integral, but we will be working with discretely layered wind data. From the compensated phase on the aperture, we can obtain the final image's intensity distribution with a Fourier transform by assuming the telescope focus operates in a Fraunhofer diffraction limit, so that electric field distribution in the image plane is the Fourier transform of the aperture function\cite{Hecht2002}.

\begin{equation}
I(\vec{r}) = | \langle \mathcal{F}(\mathcal{A}e^{i\phi_c})\rangle |^2
\end{equation}
Here $\vec{r} = (r, \theta)$ are the coordinates in the image plane, $\mathcal{A} = 1$ inside the aperture and zero outside and the brackets denote time average. We consider a telescope of aperature diameter $D = 8$ m and we do not consider apodization at the moment.

\subsection{The NOAA GFS as an atmospheric model for Cerro Pachon}
\begin{figure}[b!]
\includegraphics[width=\textwidth]{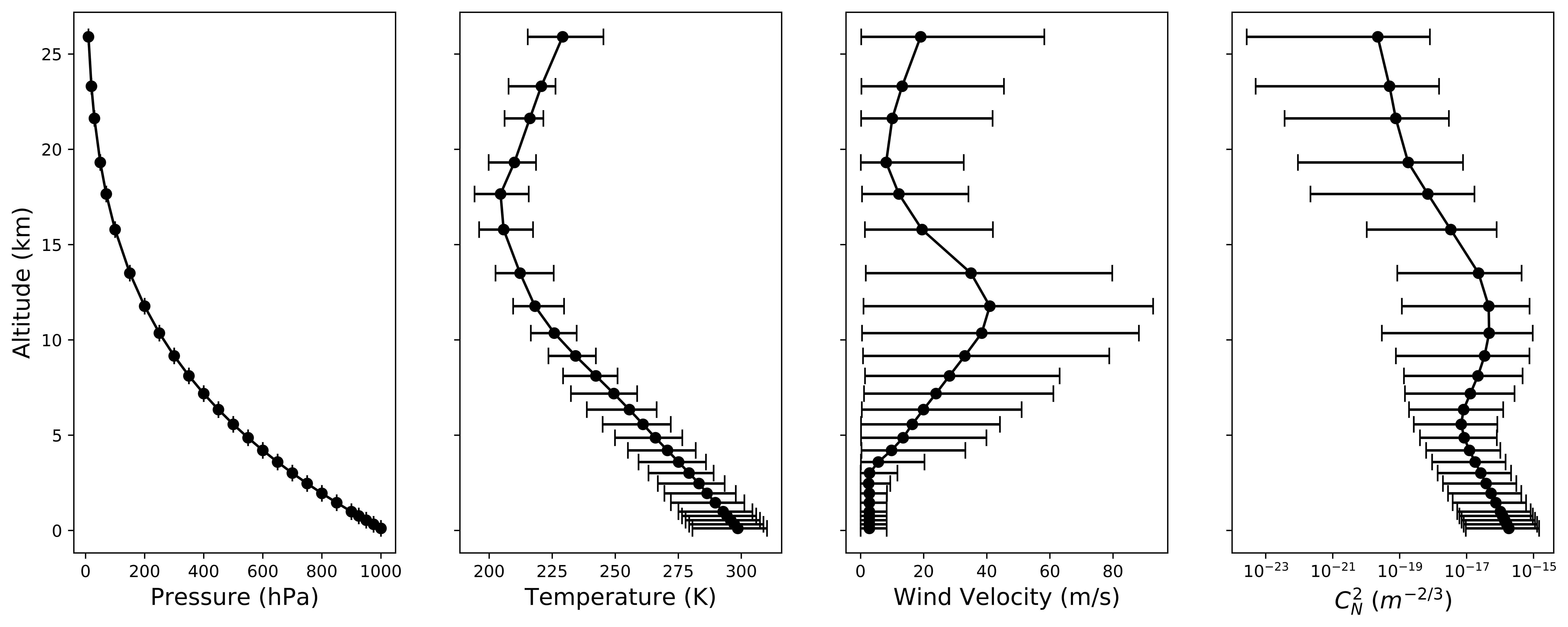}
\centering
\caption{Atmospheric Model and Turbulence Profile from NOAA GFS at Cerro Pachon assuming Hufnagel model. Due to the gamma distributed nature of the wind velocities, the error bars represent the minimum and maximum values found in each particular bin, while the dots represent the mean.}
\end{figure}

The National Oceanic and Atmospheric Organization's (NOAA) Global Forecast System (GFS) is a weather model developed by the National Centers for Environmental Prediction (NCEP), and for the considerations of the paper will be considered the atmospheric truth. The dataset is distributed as gridded data over the entire globe, will half degree scale spatial resolution, and a temporal resolution of every six hours. The dataset contains a significant number of atmospheric variables, although we will be primarily interested in the U (East) and V (North) components of wind velocities at various altitudes for construction of the $C_N^2$ profile, as well as the temperature as a function of pressure altitude to estimate the baseline index of refraction.

Because pressure variations are equalized at the speed of sound, it is common for an atmospheric model to assume that various altitudes are isobaric, to the point where pressure is the actual coordinate value in the z-direction. It is possible to formulate a rough model which relates the pressure and altitude by assuming the Earth is a uniform sphere with uniform surface temperature $T$ with an atmosphere of average molecular mass $M$. The pressure at altitude $z$ is then
\begin{equation}
\rho(z) = \rho_0 e^{\frac{-Mgz}{RT}}
\end{equation}
where $\rho_0$ is the atmospheric pressure at sea level, $g$ is the local surface gravity, and $R$ is the ideal gas constant. Using this conversion between atmospheric pressure coordinates to physical altitude, we can use the data from the NOAA GFS to estimate the index of refraction structure constant $C_N^2$. Applying a Hufnagel turbulence model\cite{Hardy1998}, we use the following equation to convert wind velocities into turbulence strength.
\begin{equation}
C_N^2(z) = e\Big[2.2\times10^{-53}z^{10}\Big(\frac{v(z)}{27}\Big)^2e^{-z/1000} + 1\times10^{-16}e^{-z/1500}\Big]
\end{equation}

Considering the Gemini South Observatory at Cerro Pachon's Coordinates of latitude -30:14:26.700 and longitude -70:44:12.096, we select the gridded bin in the GFS centered on latitude 30$\degree$ South and 70.5$\degree$ West, from December 7th 2015, to May 17th 2018, dates for which are relevant to the GPI Exoplanet Survey, and calculate the turbulence profile from the wind velocities over the range of altitudes available. The resulting turbulence profile, along with other atmospheric variables is plotted in Figure 1.

As a method of verification for this turbulence profile, we calculate the value of the Fried parameter $r_0$ at $\lambda = .5 \mu$m with\cite{Hardy1998}
\begin{equation}
r_0 = \Big[0.423k^2\int_0^{z_{\textrm{max}}}C_N^2(z)\textrm{d}z\Big]^{-3/5}
\end{equation}
and find that it is roughly 12 cm, which is better than average. Seeing conditions are $\lambda/r_0 \approx .85$ arcsec, which are fair, but slightly worse than typical conditions at Cerro Pachon, which typically sees $r_0 = 14$ cm. This is due to our estimate integrating to altitude z=0 at sea level, rather than the altitude of the observatory.

\subsection{Simulation Results}

\bgroup
\def\arraystretch{1.1}
\setlength\tabcolsep{5pt}
\begin{table}[b] 
\begin{center}
\begin{tabular}{l l l l l l l l l l l l l l}
Parameter \\
\hline
Pressure (hPa) & 1000 & 975 & 950 & 925 & 900 & 850 & 800 & 750 & 700 & 650 & 600 & 550 & 500 \\
Altitude (km) & 0.11 &  0.32 & 0.54 & 0.76 & 0.99 & 1.46 & 1.95 & 2.46 & 3.01 & 3.59 & 4.20 & 4.86 & 5.57 \\
\hline
U wind (m/s)  &  -0.38 & -0.37 & -0.37 & -0.38 & -0.38 & -0.37 & -0.32 & 2.07 & 5.55 & 10.3 & 17.4 & 24.3 & 30.1 \\
V wind (m/s)  & -3.47& -3.47 & -3.47 & -3.47 & -3.47 & -3.47 & -3.54 & -4.20 & -6.65 & -8.34 & -8.01 & -5.86 & -3.91  \\
\hline
\hline
Pressure (hPa) & 450 & 400 & 350 & 300 & 250 & 200 & 150 & 100 & 70 & 50 & 30 & 20 & 10 \\
Altitude (km) & 6.34 &  7.18 & 8.11 & 9.16 & 10.4 & 11.8 & 13.5 & 15.8 & 17.7 & 19.3 & 21.6 & 23.3 & 25.9 \\
\hline
\rule[-1ex]{0pt}{1ex}  U wind (m/s)  & 35.2 & 40.1 & 46.7 & 53.8 & 56.3 & 59.9 & 52.5 & 18.8 & 11.8 & 13.5 & 3.60 & 2.70 & 18.0  \\
V wind (m/s)  & -2.50 & 0.50 & 8.95 & 16.6 & 22.4 & 25.6 & 15.4 & 3.19 & 0.76 & 6.61 & 1.50 & 0.60 & -1.20 \\
\hline
\end{tabular}
\end{center}
\caption{Simulation Parameters. Note the strength of the wind around 200 hPa, roughly the altitude of the jet stream.}
\end{table}
\egroup

\begin{figure}[t]
\includegraphics[width=.77\textwidth]{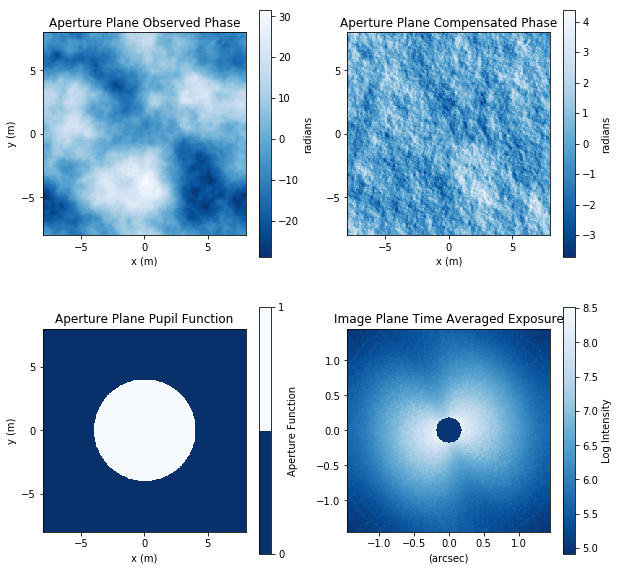}
\centering
\caption{Simulation of a Wind Butterfly. Up is North and Right is East. Top Left. Piston-subtracted phase fluctuations from the entire atmosphere over 26 Kolmogorov Layers. Top Right. Compensated phase in the aperture after a time delay $\tau = 2$ ms demonstrating the servo-lag error. Bottom Left. Pupil function for a $D = 8$ m telescope aperture without apodization. Bottom Right. Final Image, multiple summed exposures, each time step's Fourier transform in the Fraunhofer diffraction limit for the compensated phase on the top right. Black spot in the center is a crude coronagraph software mask to damp the bright central core of the PSF and bring the contrast of the butterfly shaped halo into greater visibility. Note the similarity between the pointing of the compensated phase and the dark gap in-between the butterfly wings at roughly 11 o'clock, this effect is actually orthogonal to the direction of the wind, which is pointing towards 2 o'clock.}
\end{figure}

By combining the $C_N^2$ profile we have found above with a particular set of velocities at the various altitudes encapsulated in the GFS model, we should have enough information to perform a rough model of the time evolution of the atmosphere and the response of an AO telescope to seeing such an atmosphere.

To start, L=26 unique Kolmogorov $\delta N$ screens are generated, each with a unique realization of the random noise process $\xi$, and one for each layer of the atmospheric model described earlier. Then the timestep is incremented in units of $\tau = 2$ ms, which is the time delay for an open-loop control system which roughly corresponds to the closed-loop response of GPI, which has a 3 dB bandwidth of around 20 Hz\cite{Poyneer2014}. This time delay also gives $\sim$$70$ nm RMS residual WFS error, at an atmospheric coherence time of $\tau_0 = 1.7$ ms, also roughly corresponding to models of GPI AO\cite{Poyneer16}. (As an aside, this atmospheric coherence time is unusually short, and bring into question the frozen-flow hypothesis. A more robust simulation would include boiling in the atmosphere with a clever autoregressive technique\cite{Srinath2014}, but for now we will continue to assume the frozen-flow hypothesis.) Each kolmogorov screen is translated according to the U and V components of the velocity given by the GFS model instance in Table 1, with the appropriate conversion between m/s and pixels, and with overflow causing the screen to re-enter on the opposite side, also referred to as periodic boundary conditions. At each step, the phase errors on the aperture are calculated from integrating the index of refraction and the fluctuations over the atmosphere. The compensated phase is calculated by subtracting the phase in the current timestep from that of the previous, to simulate a servo-lag error in an AO system. Then the compensated phase is Fourier transformed with a multiplicative factor for the telescope pupil to result in the final image plane. This process is demonstrated visually in Figure 2. One can see the heuristic behavior of the atmosphere and AO systems which generate the butterfly shaped halo in the PSF, and it indeed points along the direction of the wind.

\section{Butterflies in the GPIES}
Having established a semi-analytic framework and model for the origin of the wind butterfly, we will now examine its presence in the observational data taken during the GPIES campaign. Although the effect is visibly apparent and easy to spot by eye, the sheer number of images taken renders hand separation a futile task. It is with the intention of being able to distinguish images that contain wind butterflies from those who do not that we develop a metric to distinguish for us.
\subsection{Evaluation Metrics}
\begin{figure}[b]
\centering
\includegraphics[width=\textwidth]{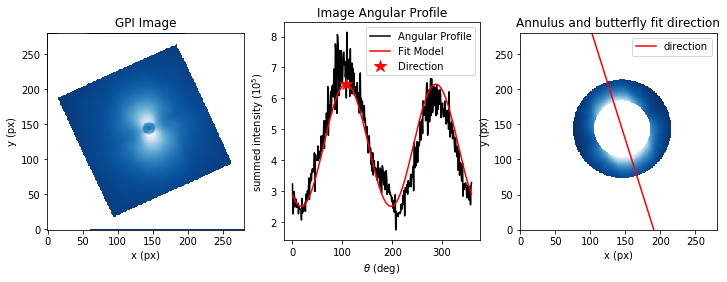}
\caption{Demonstration of extracting the preferential butterfly direction for a sample image from the GPIES campaign, as well as verifying visually that it points properly along the wind axis in the image. This particular image is rather extreme with $F=.65$.}
\end{figure}

For either simulated images, or real telescope data, there are two primary metrics of interest for evaluating the presence and strength of the wind butterfly. The first is the image angular profile, defined as
\begin{equation}
\Theta(\theta) \equiv \int_{r_{\textrm{min}}}^{r_{\textrm{max}}} I(r,\theta)\textrm{d}r
\end{equation}
Which is simply a radial integral of the image over an annulus between $r_{\textrm{min}}$ and $r_{\textrm{max}}$, to ignore complexities with coronagraphic scattering in the dark hole, and simply to examine the azimuthal symmetry or asymmetry of the halo. For azimuthal symmetry, $\Theta$ is constant, but in the presence of wind, in general, an image from adaptive optics will not be azimuthally symmetric. To first order, we can approximate the angular profile with a sinusoid of frequency $2\pi$ to fit for the preferred butterfly and wind direction.
\begin{equation}
\Theta_{\textrm{fit}}(\theta,\vec{p}) = p^{[0]}\cos(2\pi(\theta - p^{[1]})) + p^{[3]}
\end{equation}
Such a model is a function of angle in the image $\theta$ as well as a parameter vector $\vec{p}$, whose components are amplitude, phase, and a constant offset.
\begin{equation}
\vec{p} = [\textrm{amplitude},\psi, \textrm{offset}]
\end{equation}
In practice, we will only be interested in $\psi$, as it tells us of the preferred butterfly direction in the image, which is what we are interested in correlating with wind data, the other two degrees of freedom in this model simply allow the fitting to properly converge. From this model, we can construct a total error
\begin{equation}
\chi^2(\vec{p}) = \int_0^{2\pi}(\Theta - \Theta_{\textrm{fit}})^2\textrm{d}\theta
\end{equation}
which, once integrated over all angles, is a function simply of the parameter vector. We ignore normalization on $\chi^2$ at the moment since we only care about the minimum value. To obtain the minimum value of such a function, we use scipy.optimize.fmin along with a null guess for the parameters, and it converges on the optimal value for our parameter vector. An example of this fitting is demonstrated in Figure 3.

Of interest is also the fractional standard deviation in the annulus $\equiv \{ I(\vec{r}) | r_{\textrm{min}} \leq r \leq r_{\textrm{max}} \}$, defined as
\begin{equation}
F \equiv \frac{1}{\mu} \sqrt{\frac{1}{N_{\textrm{pts}}} \sum_{\textrm{annulus}} (I(\vec{r})-\mu)^2}
\end{equation}
where $\mu$ is the mean value over the annulus
\begin{alignat}{3}
\mu &=& \frac{1}{N_{\textrm{pts}}} &\sum_{\textrm{annulus}}& I(\vec{r}) \\
N_{\textrm{pts}} &=& &\sum_{\textrm{annulus}}& 1
\end{alignat}
and $N_{\textrm{pts}}$ is simply the number of discrete points or pixels under consideration for normalization. The fractional standard deviation is theoretically bounded on $[0,\infty)$, but for any reasonable and continuous image is bounded by $[0,1]$, where we consider 0 to be the case of azimuthal symmetry, and 1 to be the extremely asymmetric case of a square wave radial profile. A sine wave squared that varies from a maximum amplitude of one to zero has $F = \frac{1}{2\sqrt{2}}\approx.35$, to give a typical estimate.

\subsection{Telescope Pointing and 3D orientation of GPI Images}
The Back of the Telescope (BT) Plane is the simplest way to imagine the relationship between an image on the sky and its orientation relative to the ground. Suppose you have a DSLR on a tripod, or a multi-million dollar telescope with an Alt-Az tracking system. Either way\footnote{It is worth noting that the validity of this analogy, as well as is necessary to implement Angular Differential Imaging, a post-processing technique for combining multiple exposures while the target star moves through the zenith, that GPI operates in a fixed parallactic orientation, with the instrument derotator disabled, so that GPI is fixed with respect to the telescope orientation, which is uncommon.}, your imaging device is pointed at the celestial sphere along the line of sight vector 
\begin{equation}
\hat{r} = \langle\cos(el)\cos(az), -\cos(el)\sin(az), \sin(el)\rangle
\end{equation}
Where we have assumed the convention of the positive x-axis pointing North, and the positive y-axis pointing West. This conveniently sets up the positive z-axis to point towards Zenith, as it should. Azimuth is measured from North opening towards the East, and elevation is measured from the horizon upwards. See Figure 4 for an illustration.
With such conventions laid out, it becomes easy to identify the location of the image plane on the sky, as it must be perpendicular to the line of sight. Since there are infinitely many such planes, we will use the convention
\begin{eqnarray}
\hat{a} &=& \langle - \sin(az), - \cos(az), 0 \rangle \\
\hat{b} &=& \langle-\sin(el)\cos(az), \sin(el)\sin(az), \cos(el)\rangle
\end{eqnarray}
So that one can think of $\hat{a}$ as pointing in the direction of increasing Azimuth, and $\hat{b}$ pointing towards increasing Elevation. It is left to the reader to show that $\hat{a} \cdot \hat{b} = 0$, and that $\hat{a} \times \hat{r} = \hat{b}$ to verify the orthogonality of these unit vectors as a coordinate system.

With this elaborate set up, it becomes easy to convert vectors in the image plane into vectors in full three dimensional space, and then project them onto the ground plane. Suppose we have a wind vector which appears in the image plane rotated $\psi$ from $\hat{a}$ counterclockwise. Such a wind vector is
\begin{equation}
\hat{w} = \cos(\psi)\hat{a} + \sin(\psi)\hat{b}
\end{equation}
However, we would instead like to know $\hat{w}(\hat{x},\hat{y},{\hat{z}})$. By algebraically substituting in our coordinate vectors $\hat{a}$, and $\hat{b}$ formulas in x,y,z space, we can arrive at an expression for the wind vector in x,y,z space in terms of $\psi$, $az$, and $el$. This is
\begin{align}
\hat{w} = \langle -\cos(\psi)\sin(az) - \sin(\psi)\sin(el)\cos(az), \nonumber\\
-\cos(\psi)\cos(az) + \sin(\psi)\sin(el)\sin(az), \nonumber\\ 
\sin(\psi)\cos(el)\rangle
\end{align}
With this done, we can easily project the vector onto the ground plane by simply removing the z-component. If we need to find the direction of this wind vector as an azimuth, we can use the following trick
\begin{equation}
\textrm{azimuth} =\Big(360\degree - \arctan 2\Big(\frac{w_y}{w_x}\Big)\Big) \% 360 \degree
\end{equation}
Where $w_x$, $w_y$ are the x and y components of the wind vector, respectively, \% is the modulo operator, and it is often convenient to use a smart operator like arctan2 to get the quadrant correct.

However, images in the GPIES are not simply oriented as in the BT plane, but rather can be arbitrarily arranged due to the complexities of post-processing. Fortunately for us, the orientation of each of the image has been previously calculated in celestial coordinates. These are represented as a CD Matrix, which describe how x and y in pixels for the image correspond to right ascension and declination. Using the local sidereal time of the image during the exposure, it is possible to convert coordinates in right ascension and declination to coordinates in azimuth and elevation, using

\begin{eqnarray}
\textrm{azimuth} &=& \Big(\arctan2\Big(\frac{\cos(\delta)\sin(h)}{\sin(\phi_0)\cos(\delta)\cos(h) - \cos(\phi_0)\sin(\delta)}\Big) + 180\Big) \% 360 \\
\textrm{elevation} &=& \arcsin(\sin(\phi_0)\sin(\delta) + \cos(\phi_0)\cos(\delta)\cos(h) \\
\end{eqnarray}
where $h = \theta_L - \alpha$ is the hour angle, $\theta_L$ is the local sidereal time in radians, $\phi_0$ is the local latitude, $\alpha$ is right ascension, $\delta$ is declination, and the addition of $180\degree$ is due to the strange convention that azimuth starts from the South and opens to the West, but here we use the convention that it starts at North and opens to the East. The modulo is there to handle overflow and the azimuth and elevation are the coordinates on the sky. Once these are calculated, we can orient images relative to the BT plane because $\hat{a}$ points towards increasing azimuth and $\hat{b}$ points towards increasing elevation. Then, we use the techniques described earlier to find preferential butterfly wind vectors using our chi-squared minimization, and project them onto the ground.

\begin{figure}[b]
\centering
\begin{minipage}{.45\textwidth}
\centering
\includegraphics[width=.9\textwidth]{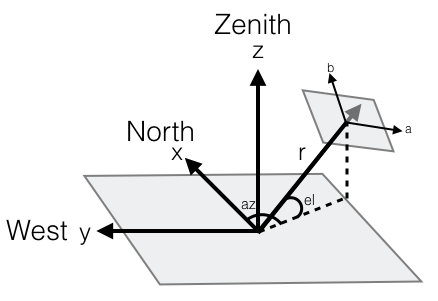}
\caption{Diagram of coordinates used to orient images on the sky demonstrating the relationship between the ground plane and the BT plane.}
\end{minipage}
\begin{minipage}{.2\textwidth}
\end{minipage}
\begin{minipage}{.45\textwidth}
\includegraphics[width=.7\textwidth]{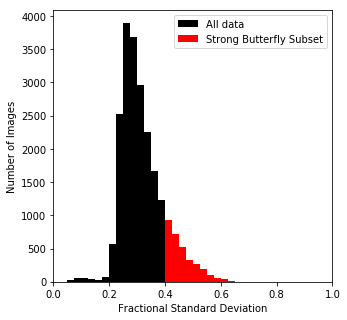}
\centering
\caption{Fractional Standard Deviation of dataset from GPIES Campaign demonstrating $1\sigma$ cutoff}
\end{minipage}
\end{figure}

\subsection{Observational Results}

\begin{figure}[b]
\includegraphics[width=\textwidth]{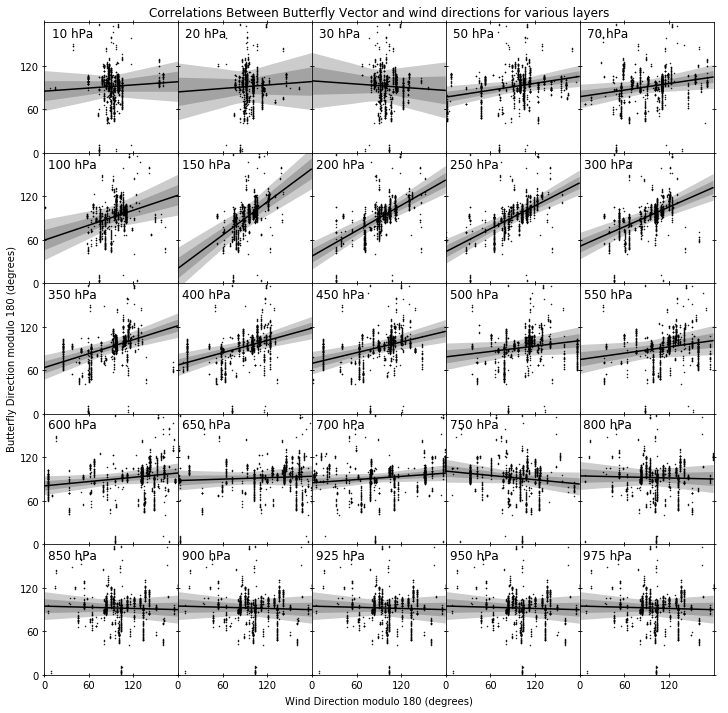}
\centering
\caption{Correlations between projected butterfly direction and wind direction both modulo 180 degrees for 25 different wind layers in the NOAA GFS. Black lines represent best-fit lines for each scatter plot, with the dark and light gray regions representing the $2\sigma$ and $4\sigma$ uncertainties in the fit line. Many wind layers correlations are negligible, with approximately horizontal fit lines implying no correlation, with the exception of the wind layers in between 100-250 hPa, which are highly correlated and nearly 1-1 in slope. These layers correspond roughly to the altitude of the jet stream, where the wind velocity and turbulence dominates the seeing effect in AO systems.}
\end{figure}

We investigate a selection of 22244 images taken during the course of GPIES campaign, and using our previously described techniques, calculate the fractional standard deviation in the images as well as determine the butterfly's wind vector and project it onto the ground coordinates. Figure 5 shows a histogram of all of the fractional standard deviations $F$ and their total number of occurrences over 40 bins. The average $F$ is around .32 and the standard deviation is around .08, and so we select a subset of very strong butterflies with $F$ greater than $1\sigma$ above average. This subset constitutes 3178 images.

The resulting dataset is matched temporally to the NOAA GFS, and correlations between the directions of the strong butterfly and the various wind layers are shown in Figure 6. Due to the availability of atmospheric data, only 1105 images are able to be compared. For each wind layer, a simple linear model is fit to the relationship between the butterfly direction and the wind layers direction, which both exist numerically as degrees azimuth in the coordinates described earlier. Slopes and their respective uncertainties are estimated to give a sense of the strength of the correlation, which should be around zero for uncorrelated, and approach one for strong correlation. The fundamental observation of this paper is the strong correlation of the butterfly and the wind at the altitudes around 100-200 hPa, or around 12-16 km, colloquially known as the jet stream. The rest of the layers exhibit either mild correlations, or strangely near the ground negative correlations, but this may be attributable to correlations in the atmospheric model from the continuity of fluid flow.

To digest the large amount of scatter plot and fit lines into a more concise form, Figure 7 was created. Both the slopes of the fit lines and the Pearson's R-coefficient defined as
\begin{equation}
R = \frac{N_{\textrm{pts}}\sum xy - (\sum x) \times (\sum y)}{\sqrt{(N_{\textrm{pts}} \sum x^2 - (\sum x)^2)\times(N_{\textrm{pts}} \sum y^2 - (\sum y)^2)}}
\end{equation}
for each scatter plot was graphed as a function of altitude. One can see both that the value of the slope and the strength of the correlation for the scatter diagrams peaks in the appropriate range of the jet stream, although the strength of the correlation, given the by the Pearson's R-coefficient, seems to peak at a lower altitude than the correspondence of the correlation, given by the slope of the best-fit line. The reason for this is unclear, but does not significantly affect our results at this time.

\begin{figure}[b]
\centering
\begin{minipage}{.45\textwidth}
\centering
\includegraphics[width=\textwidth]{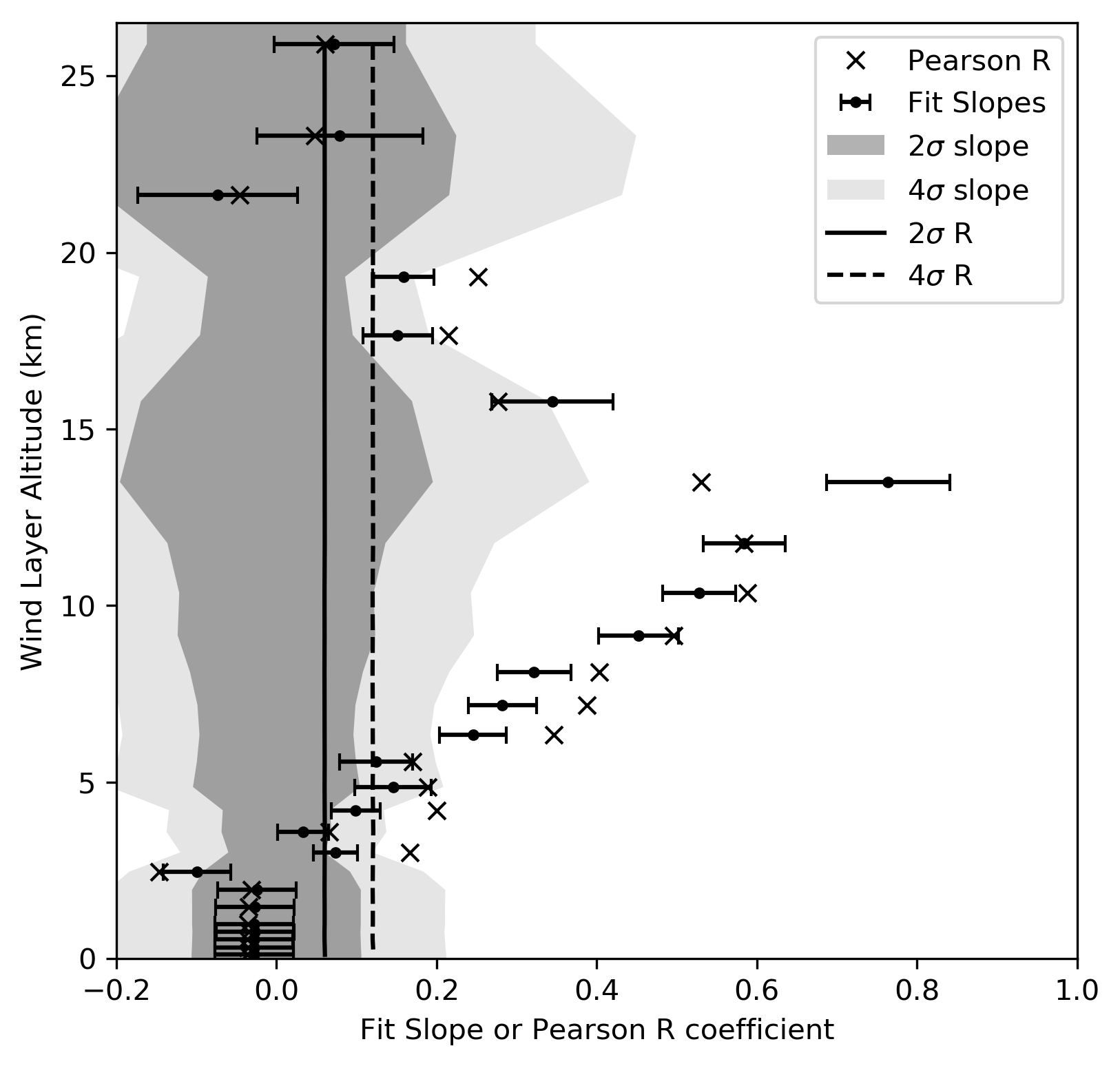}
\caption{The slopes of the best fit lines from Figure 5 over various altitudes, and error bars for the $2\sigma$ uncertainties in the fit. Also, the Person's R-coefficient for each of the scatter diagrams for comparison of the strength of the correlation to the correspondence of the directions. The filled areas and solid lines represent the bootstrapped $2\sigma$ and $4\sigma$ standard deviations for the null hypothesis.}
\end{minipage}
\begin{minipage}{.2\textwidth}
\end{minipage}
\begin{minipage}{.45\textwidth}
\includegraphics[width=\textwidth]{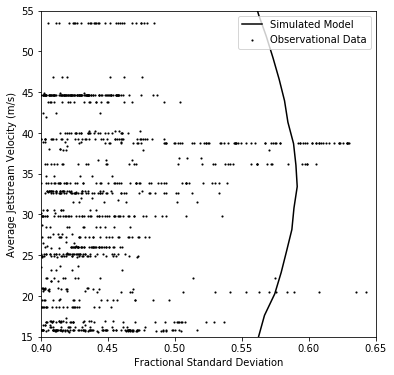}
\centering
\caption{The fractional standard deviation of the strong butterfly subset wind compared to the average velocity of the wind in the jet stream layers (100 - 300 hPa) compared to the simulated fractional standard deviations using the framework described in Section 2. Of interest is the notion that this peaks around 35-40 m/s, but drops off for both slower and faster wind speeds.}
\end{minipage}
\end{figure}

In order to quantify the likelihood of these correlations spontaneously forming due to chance, a null hypothesis of randomly distributed butterfly directions was compared to a randomly selected sample from the prior wind distributions. Our sample size of 1105 points over one hundred thousand iterations were generated in order to bootstrap an estimate for possible observed values of R or slope. The bootstrapped uncertainties in the sloped varied heavily over altitude, and the resulting areas are shown as shaded  on Figure 7. The Pearson-R formed a normal distribution with a mean of zero and standard deviation of $0.0301$ regardless of altitude, which puts the significance of our correlations at $13.2\sigma$ for $R \geq .4$, or equivalently a 1 in $10^{39}$ probability of happening due to chance.

\section{Discussion}
Although it is now clear that there is a strong relationship between the directionality of the wind butterfly and the strength of the jet stream, it is worthwhile to note another particular effect that was discovered in both the simulations and data regarding the magnitude of the the fractional standard deviation. This effect is to note that there exists a maximum value of $F$ and thus the apparent strength of the butterfly asymmetry corresponding to a particular value of the wind speed. Figure 8 demonstrate this visually.

Although there is not excellent agreement between the simulation and and the observations, one can see a similar effect in both the simulated and observed $F$ with wind velocities between 35-40 m/s in that the fractional standard deviation reaches a maximum. This makes sense when considering the analytic limits of either taking the wind velocity to infinity, or equivalently, making the AO delay infinitely slow. In this limit, the observed phase on the aperture and the correction applied by the deformable mirror will be two entirely uncorrelated Kolmogorov screens, whose difference would similarly be Kolmogorov. This would cause the resulting image to have azimuthal symmetry, as it would just be like observing through an uncorrected atmosphere. The image produced would be atmospheric speckles resolved at $\lambda/D$ forming an approximately Gaussian blob with FWHM $\lambda/r_0$. Between the peak velocity and this limit, the simulations show that the butterfly transforms into a roughly elliptically shaped halo, which should have intermediate values of $F$. In the opposite limit, taking the wind speed to zero, or equivalently taking the AO delay to zero, the correction would be perfect, with zero phase on the aperture resulting in a standard diffraction-limited Airy ring, which is also azimuthally symmetric. The very existence of the butterfly in the intermediate regime necessitates there being a particular combination of the wind velocity and AO delay corresponding to maximum azimuthal asymmetry. For our simulations, we find that this constant $v_{\textrm{peak}}\tau = 6.334 \pm 0.557$ centimeters, which corresponds to the physical distance between observed and corrected phase aberrations which would produce the greatest azimuthal asymmetry in the images. It is not known if this constant is a function of telescope diameter or observing wavelength.

The implications of this effect in terms of AO improvements is tenuous. Planet detectability algorithms often rely on the assumption that PSF noise is azimuthally symmetric, which is clearly not the case in all images, and so appropriate modeling of the asymmetry of the butterfly halo could provide improvements to detections. Improvements to AO systems response times, that is, decreasing the delay constant $\tau$ would certainly provide improvements as well, although any finite delay will still result in this image asymmetry, unless clever predictive models are developed which can anticipate atmospheric changes to apply corrections psuedo-instantaneously.

\section{Conclusion}

To conclude with a summary, we have demonstrated a semi-analytic mechanism to describe the origin on lemniscate atmospheric aberrations or colloquially, the wind butterfly, in adaptive optic telescopes, using simulations of propagating Kolmogorov turbulence and a delay to account for servo-lag errors in a real system. This demonstration motivated an exploratory data mine into images from the GPIES campaign, as well as atmospheric models. By appropriately manipulating the data, we can correlate the projected butterfly direction onto the ground with the direction of various winds layers from the atmospheric models, and we find that the wind butterfly is strongly correlated ($13.2\sigma$) with the high altitude jet-stream layers of wind around 10-15 km above the surface of the Earth. This claim reaffirms our understanding of the models that govern turbulence and adaptive optics systems, and provides an underpinning to understanding how planet detectability algorithms should handle azimuthal asymmetry in images taken during the campaign, as well as highlighting the prospective improvements that could arise from faster corrective algorithms.

\newpage
\begin{center}
\bibliography{report} 

\begin{thebibliography}{10}

\bibitem{Macintosh2014}
Macintosh, B., Graham, J.~R., Ingraham, P., Konopacky, Q., Marois, C., Perrin,
  M., Poyneer, L., Bauman, B., Barman, T., Burrows, A.~S., Cardwell, A.,
  Chilcote, J., De~Rosa, R.~J., Dillon, D., Doyon, R., Dunn, J., Erikson, D.,
  Fitzgerald, M.~P., Gavel, D., Goodsell, S., Hartung, M., Hibon, P., Kalas,
  P., Larkin, J., Maire, J., Marchis, F., Marley, M.~S., McBride, J.,
  Millar-Blanchaer, M., Morzinski, K., Norton, A., Oppenheimer, B.~R., Palmer,
  D., Patience, J., Pueyo, L., Rantakyro, F., Sadakuni, N., Saddlemyer, L.,
  Savransky, D., Serio, A., Soummer, R., Sivaramakrishnan, A., Song, I.,
  Thomas, S., Wallace, J.~K., Wiktorowicz, S., and Wolff, S., ``First light of
  the gemini planet imager,'' {\em Proceedings of the National Academy of
  Sciences}~{\bf 111}(35),  12661--12666 (2014).

\bibitem{Macintosh2015}
Macintosh, B., Graham, J.~R., Barman, T., De~Rosa, R.~J., Konopacky, Q.,
  Marley, M.~S., Marois, C., Nielsen, E.~L., Pueyo, L., Rajan, A., Rameau, J.,
  Saumon, D., Wang, J.~J., Patience, J., Ammons, M., Arriaga, P., Artigau, E.,
  Beckwith, S., Brewster, J., Bruzzone, S., Bulger, J., Burningham, B.,
  Burrows, A.~S., Chen, C., Chiang, E., Chilcote, J.~K., Dawson, R.~I., Dong,
  R., Doyon, R., Draper, Z.~H., Duch{\^e}ne, G., Esposito, T.~M., Fabrycky, D.,
  Fitzgerald, M.~P., Follette, K.~B., Fortney, J.~J., Gerard, B., Goodsell, S.,
  Greenbaum, A.~Z., Hibon, P., Hinkley, S., Cotten, T.~H., Hung, L.-W.,
  Ingraham, P., Johnson-Groh, M., Kalas, P., Lafreniere, D., Larkin, J.~E.,
  Lee, J., Line, M., Long, D., Maire, J., Marchis, F., Matthews, B.~C., Max,
  C.~E., Metchev, S., Millar-Blanchaer, M.~A., Mittal, T., Morley, C.~V.,
  Morzinski, K.~M., Murray-Clay, R., Oppenheimer, R., Palmer, D.~W., Patel, R.,
  Perrin, M.~D., Poyneer, L.~A., Rafikov, R.~R., Rantakyr{\"o}, F.~T., Rice,
  E.~L., Rojo, P., Rudy, A.~R., Ruffio, J.-B., Ruiz, M.~T., Sadakuni, N.,
  Saddlemyer, L., Salama, M., Savransky, D., Schneider, A.~C.,
  Sivaramakrishnan, A., Song, I., Soummer, R., Thomas, S., Vasisht, G.,
  Wallace, J.~K., Ward-Duong, K., Wiktorowicz, S.~J., Wolff, S.~G., and
  Zuckerman, B., ``Discovery and spectroscopy of the young jovian planet 51 eri
  b with the gemini planet imager,'' {\em Science}~{\bf 350}(6256),  64--67
  (2015).

\bibitem{Ruffio2017}
Ruffio, J.-B., Macintosh, B., Wang, J.~J., Pueyo, L., Nielsen, E.~L., Rosa, R.
  J.~D., Czekala, I., Marley, M.~S., Arriaga, P., Bailey, V.~P., Barman, T.,
  Bulger, J., Chilcote, J., Cotten, T., Doyon, R., Duchêne, G., Fitzgerald,
  M.~P., Follette, K.~B., Gerard, B.~L., Goodsell, S.~J., Graham, J.~R.,
  Greenbaum, A.~Z., Hibon, P., Hung, L.-W., Ingraham, P., Kalas, P., Konopacky,
  Q., Larkin, J.~E., Maire, J., Marchis, F., Marois, C., Metchev, S.,
  Millar-Blanchaer, M.~A., Morzinski, K.~M., Oppenheimer, R., Palmer, D.,
  Patience, J., Perrin, M., Poyneer, L., Rajan, A., Rameau, J., Rantakyrö,
  F.~T., Savransky, D., Schneider, A.~C., Sivaramakrishnan, A., Song, I.,
  Soummer, R., Thomas, S., Wallace, J.~K., Ward-Duong, K., Wiktorowicz, S., and
  Wolff, S., ``Improving and assessing planet sensitivity of the gpi exoplanet
  survey with a forward model matched filter,'' {\em The Astrophysical
  Journal}~{\bf 842}(1),  14 (2017).

\bibitem{Poyneer2016}
Poyneer, L.~A., Palmer, D.~W., Macintosh, B.~A., Savransky, D., Sadakuni, N.,
  Thomas, S.~J., V{\'{e}}ran, J.-P., Follette, K.~B., Greenbaum, A.~Z., {Mark
  Ammons}, S., Bailey, V.~P., Bauman, B., Cardwell, A., Dillon, D., Gavel, D.,
  Hartung, M., Hibon, P., Perrin, M.~D., Rantakyr{\"{o}}, F.~T.,
  Sivaramakrishnan, A., and Wang, J.~J., ``{Performance of the Gemini Planet
  Imager's adaptive optics system},'' {\em Applied Optics}~{\bf 55},  323 (jan
  2016).

\bibitem{MalesGuyon17}
Males, J.~R. and Guyon, O., ``Ground-based adaptive optics coronagraphic
  performance under closed-loop predictive control,'' {\em Journal of
  Astronomical Telescopes, Instruments, and Systems}~{\bf 4},  4 -- 4 -- 21
  (2018).

\bibitem{Rigaut1998}
Rigaut, F.~J., Veran, J.-P., and Lai, O., ``Analytical model for
  shack-hartmann-based adaptive optics systems,'' {\em Proc. SPIE} {\bf 3353},
  3353 -- 3353 -- 11 (1998).

\bibitem{Tatarski1961}
Tatarski, V.,  [{\em Wave Propagation in a Turbulent
  Medium}{\nolinebreak\hspace{0.1em}]}, Dover books on physics and mathematical
  physics, Dover (1961).

\bibitem{Johansson1994}
Johansson, E.~M. and Gavel, D.~T., ``Simulation of stellar speckle imaging,''
  {\em Proc. SPIE} {\bf 2200},  2200 -- 2200 -- 12 (1994).

\bibitem{Hardy1998}
Hardy, J.,  [{\em Adaptive Optics for Astronomical
  Telescopes}{\nolinebreak\hspace{0.1em}]}, Oxford Series in Optical \& Ima,
  Oxford University Press (1998).

\bibitem{Hecht2002}
Hecht, E.,  [{\em Optics}{\nolinebreak\hspace{0.1em}]}, Addison-Wesley (2002).

\bibitem{Poyneer2014}
Poyneer, L.~A., Rosa, R. J.~D., Macintosh, B., Palmer, D.~W., Perrin, M.~D.,
  Sadakuni, N., Savransky, D., Bauman, B., Cardwell, A., Chilcote, J.~K.,
  Dillon, D., Gavel, D., Goodsell, S.~J., Hartung, M., Hibon, P., Rantakyrö,
  F.~T., Thomas, S., and Veran, J.-P., ``On-sky performance during verification
  and commissioning of the gemini planet imager's adaptive optics system,''
  {\em Proc. SPIE} {\bf 9148},  9148 -- 9148 -- 15 (2014).

\bibitem{Poyneer16}
Poyneer, L.~A., Palmer, D.~W., Macintosh, B., Savransky, D., Sadakuni, N.,
  Thomas, S., V\'{e}ran, J.-P., Follette, K.~B., Greenbaum, A.~Z., Ammons,
  S.~M., Bailey, V.~P., Bauman, B., Cardwell, A., Dillon, D., Gavel, D.,
  Hartung, M., Hibon, P., Perrin, M.~D., Rantakyr\"{o}, F.~T.,
  Sivaramakrishnan, A., and Wang, J.~J., ``Performance of the gemini planet
  imager's adaptive optics system,'' {\em Appl. Opt.}~{\bf 55},  323--340 (Jan
  2016).

\bibitem{Srinath2014}
Srinath, S., Poyneer, L.~A., Rudy, A.~R., and Ammons, S.~M., ``Remembrance of
  phases past: An autoregressive method for generating realistic atmospheres in
  simulations,'' {\em Proc. SPIE} {\bf 9148},  9148 -- 9148 -- 8 (2014).

\end{thebibliography}
\bibliographystyle{spiebib}
\end{center}

\end{document}